\documentclass[final,5p,times,twocolumn]{elsarticle}
\usepackage{amsfonts}
\usepackage{amssymb}
\usepackage{graphics}
\usepackage{graphicx}
\usepackage{lineno}
\usepackage{times}
\usepackage{subfig}
\usepackage{hyperref}
\usepackage{natbib}

\journal{NIM A RICAP-2013}

\begin{document}
\begin{frontmatter}
\label{firstpage}

\title{Propagation of UHECRs in the Universe}
\author[uhh]{Rafael Alves Batista}
\ead{rafael.alves.batista@desy.de}
\author[uhh]{Peter Schiffer}
\author[uhh]{G\"unter Sigl}
\address[uhh]{II. Institut f\"ur Theoretische Physik - Universit\"at Hamburg, Luruper Chaussee 149, D-22761, Hamburg, Germany}


\begin{abstract}
The origin, propagation, and mechanisms of acceleration of the ultra-high energy cosmic rays (UHECRs) are not yet well understood. Aiming for a better interpretation of the available experimental data, these data have to be confronted with theoretical models. A realistic simulation of the propagation of UHECRs in the universe should take into account all the relevant energy loss processes due to the interaction with astrophysical backgrounds, as well as the intervening cosmic magnetic fields. Cosmological effects, such as the redshift dependence of the photon backgrounds and the adiabatic expansion of the universe can play an important role in the aforementioned processes. Here we present results of simulations of the propagation of UHECR through the large scale structure of the universe considering cosmological and magnetic field effects simultaneously.
\end{abstract}


\end{frontmatter}

\section{Introduction}

During their propagation to Earth, ultra-high energy cosmic rays (UHECRs) can suffer different interaction processes and be deflected by the pervasive cosmic magnetic fields, which directly affects the spectrum and chemical composition observables of these particles. Therefore, to address the question concerning the origin and nature of UHECRs, one has to simulate their propagation taking into account all the relevant interaction and energy loss processes, as well as the effects of cosmic magnetic fields. 

If UHECRs have an extragalactic origin, the intervening cosmic magnetic fields can play an important role on propagation, and increase the trajectory length of the particles in such a way that cosmological effects, such as the adiabatic expansion of the universe and the redshift evolution of the photon backgrounds, can have a relevant contribution to the energy losses. Weak magnetic fields should only slightly increase the trajectory length for nearby sources, but should have considerable effects for distant sources. Depending on the model of extragalactic magnetic field (EGMF) used this increase in path length causes larger deflections. By propagating UHECRs through the large scale structure of the universe (LSS) obtained from magnetohydrodynamical (MHD) simulations, Dolag {\it et al.}~\cite{dolag04} concluded that even for heavy nuclei the expected deflection due to the cosmic magnetic fields are of the order of a few degrees, whereas Sigl {\it et al.}~\cite{sigl04} concluded that for protons the deflections can be small in some specific regions of the sky, but are overall larger for heavier nuclei. These results are crucial to understand whether it is possible to do astronomy with UHECRs.

Aiming for an interpretation of the experimental data, it is possible to simulate the propagation of UHE particles and check whether the theoretical predictions match the observations. The main observables that propitiate this comparison are the spectrum, composition and anisotropy. 

The shape of the spectrum is not affected by magnetic fields if the sources are identical and uniformly distributed, with separation distance much less than the characteristic propagation lengths, as stated by the propagation theorem~\cite{aloisio04}. If these conditions are satisfied, then the spectrum has a universal form. 

The chemical composition of UHECRs is still unknown, and the experimental data hint towards contradictory results. The Pierre Auger Collaboration reported a gradual increase in mass number towards higher energies~\cite{auger10}, whereas results from HiRes are compatible with a proton composition at the highest energies~\cite{hires10}. Therefore, when searching for realistic scenarios of propagation of UHECRs the mass composition also has to be explained, together with the spectrum.

The third important observable is the anisotropy. In this sense, a realistic scenario for UHECR propagation should explain, apart from the spectrum and composition, also the observed (an)isotropy. However, this observable is not robust, since it is directly affected by the galactic magnetic field (GMF). In the recent GMF model proposed by Jansson and Farrar~\cite{jansson12a,jansson12b} the average deflection is strongly non-uniform across the sky, and can be $\sim$5.2$^\circ$ for 60 EeV protons. One should also add the deflections induced by the EGMF, and hence the combination of galactic and extragalactic deflections can strongly affect the anisotropy observable.

In the present work we use the same LSS simulation used by Sigl {\it et al.}~\cite{sigl04}. To propagate the UHECRs through the universe we have used the CRPropa 3 code~\cite{alvesbatista13,kampert13,armengaud07}. In this code there are two different modes of propagation available, namely a unidimensional (1D) and a tridimensional case (3D). The redshift evolution of the backgrounds and energy losses due to the adiabatic expansion of the universe are taken into account in the 1D  case, but not in the 3D case. The reason for this is that if magnetic fields are considered in a 3D simulation, the information regarding the effective trajectory length of the particles, and therefore the redshift, is not known beforehand. In the present work we address this problem by using an approximate method to correct the 3D simulations taking into account redshift losses. 

This work is divided as follows: in section \ref{sec:propagation} we describe the main energy loss processes that take place at UHE; in section \ref{sec:method} we present a method used to account for cosmological effects in 3D simulations; in section \ref{sec:applications} we apply this method to simulated data; finally, in section \ref{sec:conclusions} we summarize the work and present the final remarks and futures perspectives.

\section{Energy losses and interactions}\label{sec:propagation}

UHECRs lose energy during the propagation to Earth due to four main processes: pair production, pion production, and nuclear decay/photodisintegration (in the case of nuclei) and adiabatic expansion of the universe. The pair production is a highly inelastic process and causes a continuous energy loss that can be analytically computed. The photopion production interactions for nuclei can be numerically treated with the SOPHIA code~\cite{mucke00}. Nuclear decay comprises $\alpha$ and $\beta$ decays, as well as nucleon dripping. Photodisintegration is the dominant energy loss process that takes place at ultra-high energies for nuclei heavier than hydrogen. The adiabatic expansion of the universe is also a source of energy loss and is relevant for particles that transversed cosmological distances. The cosmological framework used here assumes a flat universe with matter and dark energy only, ignoring the radiation component. This assumption is valid for $z \lesssim 10^3$. The matter and dark energy densities are respectively $\Omega_m=0.3183$ and $\Omega_\Lambda=0.6817$, and the Hubble constant is $H_0=67.04$ km/s/Mpc.

The two main photon fields that should be taken into account in the propagation of UHECRs in the universe are the CMB and the CIB. The treatment of the CMB is straightforward, since it can be done analytically and its redshift evolution is given by a factor $(1+z)^{2}$ with respect to the photon density of this background at $z=0$. The CIB has to be treated numerically, but its redshift evolution is assumed to be the same as the CMB within the framework of CRPropa.

\section{Accounting for cosmological effects and magnetic fields simultenously} \label{sec:method}

As discussed before, the redshift dependence of the photon backgrounds, namely the CMB and CIB, and the evolution of the sources can affect the spectrum and composition of the detected cosmic rays. Also, the expansion of the universe causes energy losses.

From a 3D simulation one can obtain the following quantities: initial energy of the particle ($E_i^{3D}$), final energy of the particle ($E_f^{3D}$), initial $(A_i^{3D},Z_i^{3D})$ and final $(A_f^{3D},Z_f^{3D})$ particle type, initial particle position $(x_i,y_i,z_i)$ and final particle position $(x_f,y_f,z_f)$, effective propagation time ($T^{3D}$), among others.

To introduce redshift effects in the 3D simulation, we resimulate each individual particle in 1D, using the following input from the 3D simulation: $E_i^{3D}$, $(A_i^{3D},Z_i^{3D})$ and $T^{3D}$. For each resimulated particle we obtain a set of subproducts from the photodisintegration of the mother-nucleus. Let $k$ be an index associated to each one of these subproducts. Then the output of the 1D simulation provides us the final energy $E_{f,k}^{1D}$ and the final particle type $(A_{f,k}^{1D},Z_{f,k}^{1D})$. As a first approximation we can choose $k$ uniformly among all daughter-nuclei. The next step of this method is to replace the final energy and type of the particle in the 3D simulation by the one obtained from the 1D simulation, by doing  $E_f^{3D}=E_{f,k}^{1D}$ and $(A_f^{3D},Z_f^{3D})=(A_{f,k}^{1D},Z_{f,k}^{1D})$.

\begin{figure}[!h]
    \includegraphics[width=0.98\columnwidth]{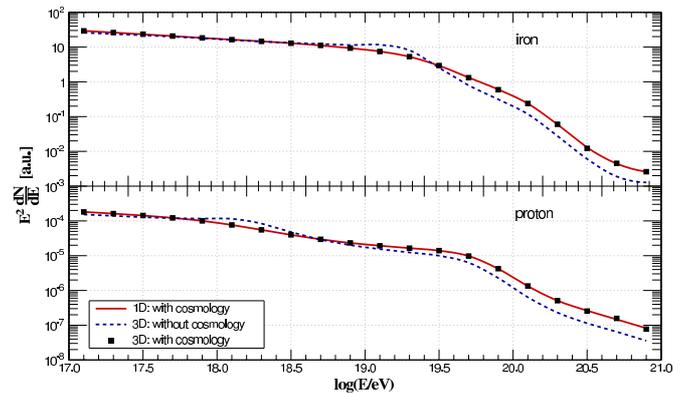}
    \caption{Spectra for the 1D simulation (solid line), the 3D without cosmology (dashed line) and the 3D corrected for cosmology (squares) for the a pure iron (above) and pure proton (below) composition. The spectra are normalized to the total number of events times 100 for iron, and 0.01 for proton.}
    \label{fig:speccheck}
\end{figure}

To check this method, we have applied it to the trivial case of a 3D simulation without magnetic fields. The sources were assumed to be uniformly distributed up to 4000 Mpc and to have a spectral index of -2.2. In the first case we considered only protons, whereas in the second we have considered only iron. The parameters of the 1D simulation are exactly the same, so that the only difference between the 1D and 3D cases are the cosmological effects. The simulated spectra in 1D and 3D, as well as the corrected one, can be seen in figure \ref{fig:speccheck} for a pure iron (upper) and pure proton (lower) composition. 

\section{Applications} \label{sec:applications}

The simulations presented here assumed the source distribution from Ref.~\cite{sigl04} according to a differential energy spectrum that follows a power law of spectral index $\gamma=-1$, reweighted afterwards to $\gamma=-2.2$. For the 3D case the observer, assumed to be a sphere of radius 0.5 Mpc, lies within a cubic box concatenated with several copies of itself up to 2 Gpc, assuming periodic boundary conditions. The propagation is performed by integrating the Lorentz force using an integrator with adaptative step sizes in the range between 10 kpc and 1 Mpc. The relevant interactions, namely pion production, photodisintegration, pair production and redshift losses (only for the 1D case) were used. The astrophysical backgrounds considered were the CMB and CIB. The model by Kneiske {\it et al.}~\cite{kneiske04} was used to treat the interactions with CIB. The redshift evolution of the photon fields was  also done according to this reference. 

\begin{figure}[!h]
    \includegraphics[width=0.98\columnwidth]{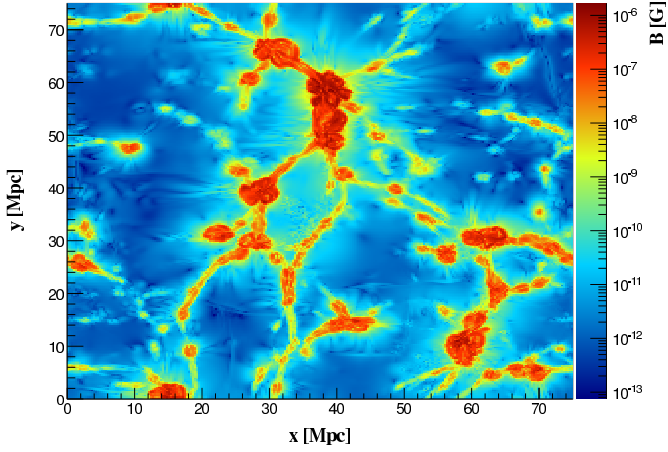}
    \caption{Two-dimensional cut through the magnetic field box. The color scale indicates the strength of this field.}
    \label{fig:MiniatiField}
\end{figure}

To understand the effects of cosmic magnetic fields on the spectrum, we have performed simulations with and without the magnetic field from Ref.~\cite{sigl04}. A side view of this magnetic field box is shown in figure~\ref{fig:MiniatiField}. All the simulations were performed for two mass compositions, namely pure iron and pure proton. 

\begin{figure}[!h]
    \includegraphics[width=0.98\columnwidth]{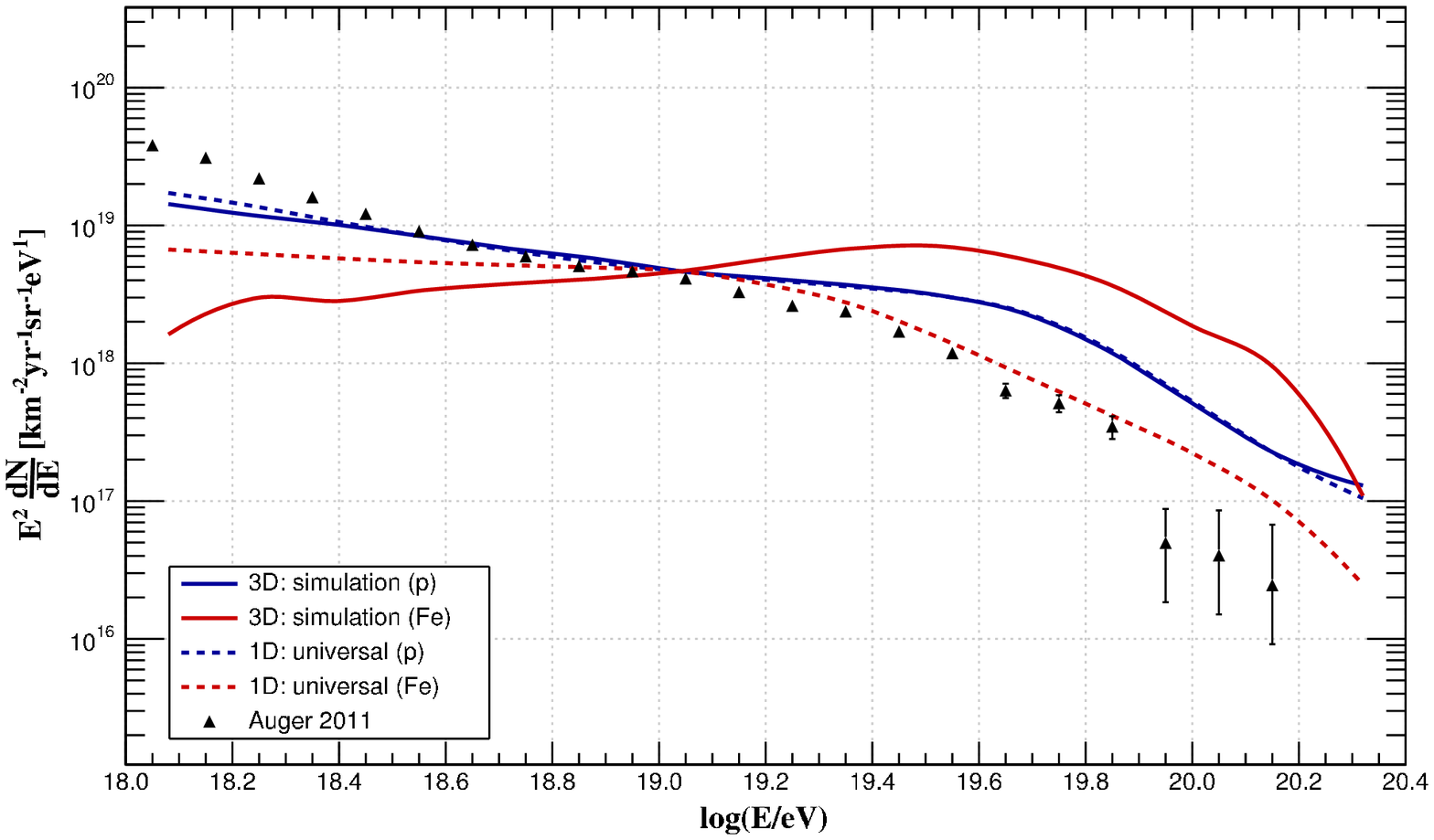}
    \caption{Simulated spectra corrected for cosmology (solid lines), universal spectrum (dashed lines) for a proton (blue) and iron (red) compositions. The dots correspond to the hybrid Auger spectrum, taken from Ref. \cite{settimo12}. The spectra here presented are normalized to the Auger spectrum at $10^{19.1}$ eV.}
    \label{fig:simspectra}
\end{figure}

The results of these simulations are shown in figure \ref{fig:simspectra}, together with the corresponding universal spectra and the experimental spectrum measured by the Pierre Auger Observatory~\cite{settimo12}, for the sake of comparison.

\begin{figure}[!h]
    \includegraphics[width=0.98\columnwidth]{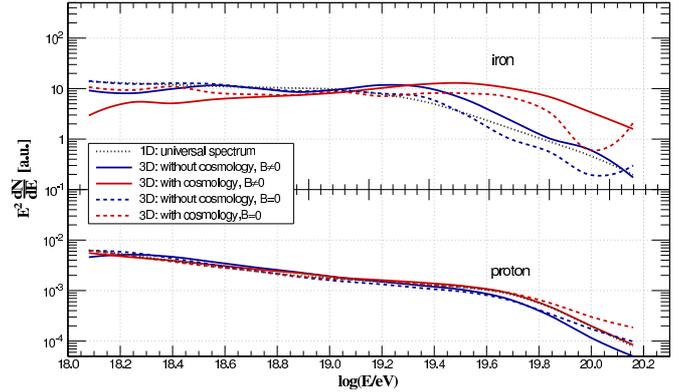}
    \caption{Simulated spectra corrected for cosmology (red lines), not corrected (blue lines), for the case with (solid lines) and without (dashed lines) magnetic field. The universal spectrum is represented by the black dotted lines, for comparison. The spectra are normalized to the total number of events times a factor of 100 for the iron case, and 0.01 for the proton case.}
    \label{fig:spectracomparison}
\end{figure}

If the source distribution is not uniform, then the previously described propagation theorem does not predict a universal spectrum, and therefore a difference between the spectrum in the cases with and without magnetic field is expected. This can be seen in figure \ref{fig:spectracomparison}. The effects of magnetic fields on the spectrum are much larger in the iron than in the proton scenario, since the charge of the cosmic ray produced at the source in the former case is 26 times the proton charge, thus implying deflection angles between 1 and 26 times larger, due to the containment of these particles around the source.

In figure \ref{fig:spectracomparison} it is also possible to compare the original 3D spectrum with the one corrected for cosmology. There are visible effects of cosmology for both simulated mass compositions scenarios. This happens because when the cosmology correction is applied, many events from the lower energy tail of the spectrum will suffer another energy loss process not present in the original 3D simulation, and will drop below the minimum energy threshold, set to 1 EeV in our simulation. 


\begin{figure}[!h]
    \includegraphics[width=0.98\columnwidth]{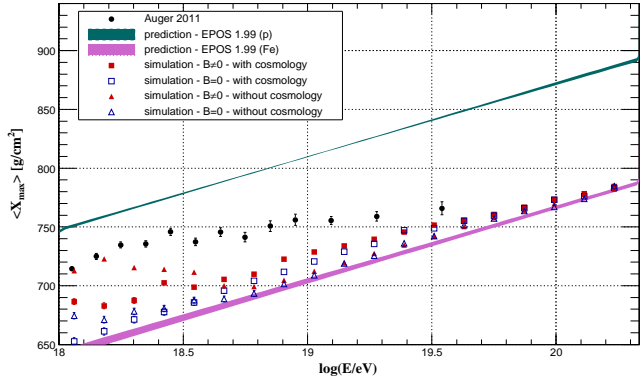}
    \caption{Mean value of the $\left<X_{max}\right>$ for the simulated data, compared to measurements of the Pierre Auger Observatory (black circles) and the theoretical predictions according to the EPOS 1.99 model (filled area) for protons (green filled area) and iron (pink filled area). The scenarios without magnetic field  are represented by empty markers, whereas the scenarios with magnetic fields are represented by filled markers. Squares correspond to the simulated data set with the cosmology correction, and the triangles without it. }
    \label{fig:composition}
\end{figure}

We are also interested in the change in mass composition due to magnetic fields and cosmology effects. This is analyzed only in the pure iron scenarios, since the proton case is trivial. The mean value of the depth of the shower maximum ($\left<X_{max}\right>$) using the parametrization from Ref.~\cite{auger13} is used as an observable for the mass composition. The value of $\left<X_{max}\right>$ for different energy ranges are shown in figure \ref{fig:composition}, together with the experimental values measured by the Pierre Auger Observatory~\cite{auger13} and the theoretical predictions assuming the EPOS 1.99 hadronic interaction model.

In this figure it can be seen that even though cosmological effects are relevant, magnetic fields are responsible for the largest change in composition at lower energies.

\begin{figure}[!h]
    \includegraphics[width=0.98\columnwidth]{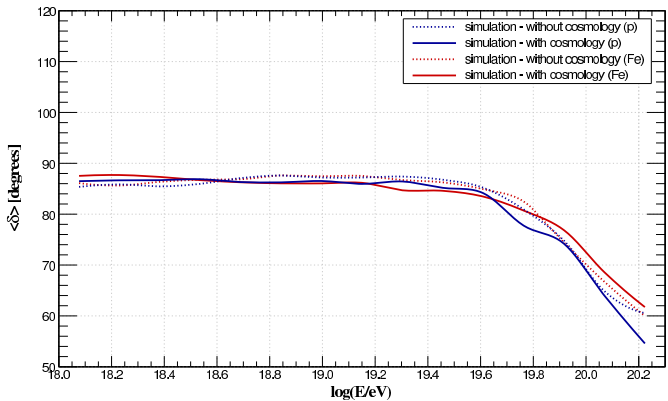}
    \caption{Mean value of the deflection as a function of the energy for an iron (red) and proton (blue) composition. The dotted lines correspond to the original 3D simulation, whereas the solid lines correspond to the simulation including cosmological effects.}
    \label{fig:aniso}
\end{figure}

In figure~\ref{fig:aniso} the mean deflection as a function of the energy is shown. The differences in the energy dependence of the deflection due to magnetic fields and cosmology are small. In case of isotropy, the mean value of the deflection would be 90$^\circ$. We notice that for lower energies the simulated data sets are highly isotropic, becoming more anisotropic at higher energies.

\section{Conclusions} \label{sec:conclusions}

In this work we have presented a new method to account for cosmological effects in 3D simulations of propagation of UHECRs. This method allows us to perform 3D simulations of the propagation of UHECRs, including heavy nuclei, considering all relevant energy loss and interaction processes, as well as the expected magnetic field and baryon density obtained from MHD simulations.

We applied this method to the same MHD simulation used by Sigl {\it et al.}~\cite{sigl04} for a scenario with a pure proton composition, and another one with pure iron. We have shown the effects of cosmology and magnetic fields, separately, upon the spectrum and observed mass composition observables. From the plots previously presented it is clear that a 3D simulation is required, since magnetic fields affect the shape of the spectrum. It is also clear that cosmological effects play an important role in the spectrum and mass composition observables.

Our results have strong implications for cosmic ray astronomy. In our particular example the deflections are so large that it is impossible to trace back the sources, especially if the fraction of heavier nuclei is large. However, the results of this work are bounded by the baryon density and magnetic fields obtained from a non-constrained MHD simulation, which was done assuming a given set of initial conditions. The effects of these assumptions on the propagation of UHECRs are not totally known. Future studies using new and constrained MHD simulations are currently underway and may favor or disfavor the results here presented regarding the possibility of doing cosmic ray astronomy.

\section*{Acknowledgements}

This work was supported by the Deutsche Forschungsgemeinschaft through the collaborative research centre SFB 676, by BMBF under the grant 05A11GU1, by the Helmholtz Alliance for Astroparticle Phyics (HAP) funded by the Initiative and Networking Fund of the Helmholtz Association, and by the Forschungs- und Wissenschaftsstiftung Hamburg through the program ``Astroparticle Physics with Multiple Messengers''.

\end{document}